\documentclass{Interspeech}
\interspeechcameraready 
\title{DS-Codec: Dual-Stage Training with Mirror-to-NonMirror Architecture Switching for Speech Codec\thanks{$^{*}$Corresponding author}}

\author[affiliation={1}]{Peijie}{Chen}
\author[affiliation={2}]{Wenhao}{Guan}
\author[affiliation={1}]{Kaidi}{Wang}
\author[affiliation={1}]{Weijie}{Wu}
\author[affiliation={1}]{Hukai}{Huang}
\author[affiliation={1}]{Qingyang}{Hong$^{*}$}
\author[affiliation={2}]{Lin}{Li$^{*}$}

\affiliation{School of Informatics}{Xiamen University}{China}
\affiliation{School of Electronic Science and Engineering}{Xiamen University}{China}
\email {peijiechen@stu.xmu.edu.cn}


\usepackage{comment}
\usepackage{multirow}
\usepackage{algorithm}
\usepackage{algpseudocode}
\usepackage{svg}

\begin{document}

\maketitle

\begin{abstract}
    Neural speech codecs are essential for advancing text-to-speech (TTS) systems. With the recent success of large language models in text generation, developing high-quality speech tokenizers has become increasingly important. This paper introduces DS-Codec, a novel neural speech codec featuring a dual-stage training framework with mirror and non-mirror architectures switching, designed to achieve superior speech reconstruction. We conduct extensive experiments and ablation studies to evaluate the effectiveness of our training strategy and compare the performance of the two architectures. Our results show that the mirrored structure significantly enhances the robustness of the learned codebooks, and the training strategy balances the advantages between mirrored and non-mirrored structures, leading to improved high-fidelity speech reconstruction.
\end{abstract}

\section{Introduction}
In recent years, large language models (LLMs) \cite{achiam2023gpt,hurst2024gpt} have exhibited remarkable capabilities in the realm of text generation, garnering significant attention across various domains, including the field of text-to-speech synthesis (TTS) \cite{wang2023neural,borsos2023audiolm,kreuk2022audiogen,10889258}. A pivotal challenge is the transformation of continuous speech signals into interpretable representations suitable for inference and training within large language models. Consequently, developing a high-quality speech tokenizer \cite{baevski2020wav2vec,schneider2019wav2vec,hsu2021hubert} emerges as a critical component in advancing speech synthesis technologies. Neural speech codecs, a prominent category within this sphere, aim to efficiently compress speech signals into lossy discrete representations at reduced bitrates, striving to retain the maximal amount of speech information. This pursuit is integral to the seamless integration of speech processing with the robust frameworks of large language models.

Most speech codec models typically comprise three integral components \cite{van2017neural,zeghidour2021soundstream}: the encoder, the quantization module, and the mirrored decoder. The encoder is responsible for encoding the speech signal into the latent representations. Subsequently, the quantization module discretizes these latent representations into discrete tokens \cite{vasuki2006review}. Predominantly, the quantization process employs the Residual Vector Quantization (RVQ) methodology \cite{kumar2024high,defossez2022highfi,yang2023hifi}, which is widely recognized for its efficacy in balancing compression efficiency with the preservation of speech quality. The mirrored decoder then reconstructs the speech signal from these discrete tokens, aiming to maintain high fidelity in the synthesized speech output. These components are fundamental to the functionality of neural speech codecs, enabling efficient compression and reconstruction of speech signals.

However, conventional RVQ-based speech codecs typically necessitate multiple token sequences to represent speech, which is incongruent with the single-sequence input paradigm favored by large language models. This discrepancy may require an additional non-autoregressive (NAR) stage \cite{wang2023neural} to handle the supplementary codebook token sequences effectively. In recent developments, the domain of single-codebook neural speech codecs has witnessed rapid advancement \cite{ji2024wavtokenizer,xin2024bigcodec,li2024single}. Numerous studies have explored various techniques to augment model performance, among which the product quantization \cite{guo2024addressing} has demonstrated considerable promise, particularly in the context of large codebooks and enhanced codebook utilization. This innovation has prompted us to investigate the viability of employing product vectors to construct expansive codebooks.

Additionally, traditional models like Encodec \cite{defossez2022highfi}, and advanced models such as DAC \cite{kumar2024high} adopt a mirrored structure. In contrast, FACodec \cite{ju2024naturalspeech} and Wavtokenizer \cite{ji2024wavtokenizer} emphasize the decoder's greater importance over the encoder in the acoustic codec reconstruction process, leading them to adopt the non-mirrored decoder upsampling structure. This architectural divergence also inspires us to re-think speech codec from a structural perspective. 

In this paper, we introduce DS-Codec, a neural speech codec based on a dual-stage training strategy. Our proposed model achieves exceptional speech reconstruction results and surpasses the performance of most single-codebook speech codecs. The contributions of this paper are as follows:

\begin{itemize}
    \item We propose a novel neural speech codec, named DS-Codec, based on vector quantization (VQ) and product quantization (PQ) separately, which employs a single codebook that avoids the extra complexity of multiple codebooks.
    \item We introduce a new training strategy with mirror and non-mirror architectures switching to balance the advantages between the two architectures, aiming to improve the robustness of the codebook and speech reconstruction performance. 
\end{itemize}

The reconstructed speech from various speech codec models are available at \href{audio sample}{https://pppjchen.github.io/DSCodec.}

\section{Methods}
\begin{figure*}[t]
  \centering
  \includegraphics[width=1\linewidth]{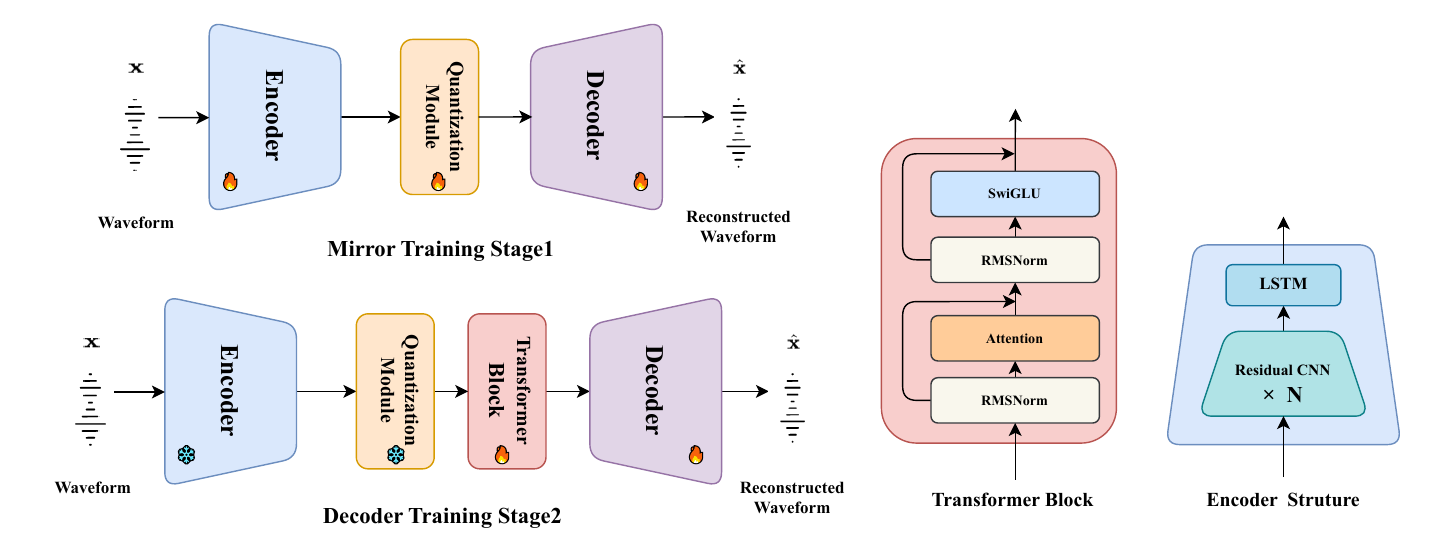}
  \caption{Schematic diagram of DS-Codec and illustration of the dual-stage training strategy for DS-Codec. The codec with non-mirror architecture is composed of a mirrored Encoder, Quantization Module, Transformer Block, and mirrored Decoder.}
  \label{fig:model}
\end{figure*}

\subsection{Model Architecture}
The overall framework of our model adopts a non-mirrored architecture, as illustrated in Figure \ref{fig:model}. 
The Encoder and Decoder architectures are inspired by BigCodec \cite{xin2024bigcodec}, with the Encoder comprising a series of residual Convolutional Neural Network (CNN) blocks. Each block incorporates snake activation functions \cite{ziyin2020neural} and is followed by a two-layer unidirectional Long Short-Term Memory (LSTM) network. The CNN blocks are designed to downsample the input waveforms by a specific factor, utilizing multiple convolutional layers with varying dilation rates to effectively capture sequential data patterns. Together, these five modules achieve a cumulative downsampling factor of 200. The Decoder mirrors the Encoder's structure, employing transpose convolutions for upsampling to reconstruct the original waveform with high fidelity. To further enhance speech reconstruction capabilities and leverage the Transformer's exceptional context-building abilities, we introduce the Transformer Block, inspired by LLaMA's \cite{touvron2023llama} decoder layer. This block integrates residual Attention and residual SwiGLU components, both enhanced with RMSNorm for normalization.

We also utilize the same discriminator architecture as BigCodec, which includes two discriminators: a multi-period discriminator (MPD) introduced in HiFiGAN \cite{kong2020hifi} and a multi-scale short-time Fourier transform (MS-STFT) discriminator used in EnCodec \cite{defossez2022highfi}. This dual-discriminator design ensures robust discrimination across both time and frequency domains, significantly improving the model's overall performance.

\subsection{Quantization Module}
\subsubsection{Vector Quantization}
The vector quantization (VQ) module builds upon the methodology proposed by Yu et al. \cite{yu2021vector}, which employs a fixed-size single-codebook containing 8,192 codes to map latent representations into discrete vectors. To optimize codebook utilization, the latent representations undergo a dimensionality reduction process before quantization: they are first projected into a low-dimensional space (with a dimension of 8, as suggested in DAC), quantized, and subsequently projected back to their original dimensionality. Additionally, both the latent variables and the codebook vectors are L2-normalized to enhance the efficiency and accuracy of the quantization process.
\subsubsection{Product Quantization}
Product Quantization (PQ) \cite{jegou2010product} employs multiple Vector Quantization (VQ) modules. The latent representations are divided along the channel dimension into non-overlapping segments, with each segment assigned to a dedicated VQ module. These segments are independently quantized by their corresponding VQ modules, and the resulting quantized vectors are concatenated to form the final PQ output $\hat{y}$. During the training and inference, it can be treated as a single codebook in the following way. As shown in the algorithm below.

\begin{algorithm}
    \caption{Product Quantization}
    \label{alg:pq}
    \renewcommand{\algorithmicrequire}{\textbf{Input:}}
    \renewcommand{\algorithmicensure}{\textbf{Output:}}
    
    \begin{algorithmic}[1]
        \Require $y = encoder(x)$ the output of the encoder, vector quantizers $Q_i$ for $i = 1,2... N_q$, codebook size $S_i$ for $i = 1,2... N_q$
        \Ensure the quantized $\hat{y}$, codebook index $\hat{code}$
        \State split $y$ into $N_q$ groups in channel dimension;
        \For{$i = 1$ to $N_q$}
            \State $\hat{y}_i,\hat{code_i} \gets Q_i(y_i)$
            \If {$i==1$}
                \State $\hat{y} = \hat{y}_i$; $\hat{code} = \hat{code_i}$
            \Else
                \State $\hat{y} = concat(\hat{y},\hat{y}_i)$; $\hat{code} = \hat{code} * S_{i-1}+\hat{code_i}$
            \EndIf
    \EndFor
    \State \textbf{return} $\hat{y},\hat{code}$
    \end{algorithmic}
\end{algorithm}
\begin{table*}[h!]
\caption{Speech reconstruction performance comparison of different models. Main results of DS-Codec on the LibriSpeech test set with 2620 utterances.}
\centering
\begin{tabular}{ccccccccc}
\toprule
\textbf{Model} & \textbf{Bandwidth $\downarrow$} & \textbf{Nq $\downarrow$} & \textbf{token/s $\downarrow$}  & \textbf{UTMOS $\uparrow$} & \textbf{PESQ $\uparrow$} & \textbf{STOI $\uparrow$} & \textbf{F1 Score $\uparrow$} \\
\midrule
    GT & - & - & - & 4.086 & - & - & - \\
    DAC & 4.0kpps & 8 & 400 & 3.325 & 2.722 & 0.938 & 0.946 \\
    Encodec & 6.0kbps & 8 & 600 & 3.074 & 2.756 & 0.938 & 0.945 \\
    Ticodec & 3.0kpps & 4 & 300 & 3.501 & 2.373 & 0.919 & 0.936 \\
\midrule
    Encodec & 1.5kbps & 2 & 150 & 1.582 & 1.560 & 0.845 & 0.836 \\
    Ticodec & 1.5kbps & 2 & 150 & 3.347 & 1.921 & 0.880 & 0.917 \\
\midrule
    DAC & 1.0kbps & 1 & 100 & 1.246 & 1.056 & 0.617 & 0.552 \\
    Ticodec & 0.75kbps & 1 & 75  & 3.052 & 1.553 & 0.832 & 0.895 \\
    Wavtokenizer & 0.9kbps & 1 & 75 & 3.784 & 2.114 & 0.897 & 0.911 \\
    Bigcodec & 1.04kbps & 1 & 80 &4.108 & 2.681 & 0.935 & 0.942 \\
\midrule
    % DS-Codec-PQ-apc & 1.28kbps & 1 & 80 & 4.164 & 2.655 & 0.934 & 0.940 \\
    DS-Codec-PQ & 1.28kbps & 1 & 80 & 4.214 & \textbf{2.882} & \textbf{0.941} & 0.943 \\
    DS-Codec-VQ & 1.04kbps & 1 & 80 & \textbf{4.218} & \textbf{2.862} & \textbf{0.941} & \textbf{0.944} \\
\bottomrule
\end{tabular}

\label{tab:performance}
\end{table*}

In this study, we create 65,536 codebook sizes based on four codebooks with sizes of [16, 16, 16, 16] in the DS-Codec-PQ, The VQ loss is the sum of the VQ losses from each VQ module.

\subsection{Training Strategy}
Inspired by APCodec+ \cite{du2024apcodec+}, we employed a novel staged training strategy, which is also a two-stage training strategy, as shown in Figure \ref{fig:model}. 

\begin{itemize}
    \item \textbf{Mirror Training: }The first stage employs a traditional joint training approach with standard mirrored speech codec frameworks. The model is trained with the Encoder, the Quantization Module, and the mirrored Decoder. During this stage, the model prioritizes the development of a robust codebook leveraging the mirrored design, enabling high-quality speech reconstruction capabilities.
    \item \textbf{Decoder Training: }In the second stage, we focus on enhancing the decoder's capability, emphasizing its critical role in generating high-quality speech within speech codecs. This stage aims to further improve the decoder's performance and overall reconstruction quality. To this end, we freeze the parameters of the encoder and quantizer, integrate the Transformer Block, switch to the non-mirrored architecture, and reinitialize the parameters of the discriminator. This ensures a more efficient and stable training process, thus improving the speech reconstruction capability. Unlike APCodec+, we avoid reinitializing the Decoder’s parameters, as it has already exhibited robust speech reconstruction capabilities during the first training stage. Instead, we retain the Decoder's parameter weights from the first stage and reduce the learning rate size. This approach accelerates the training process by minimizing unnecessary adjustments.
\end{itemize}

\section{Experiments}

\subsection{Dataset and Metrics}\label{metrics}

We use LibriSpeech \cite{panayotov2015librispeech} to train the speech codec we proposed. The LibriSpeech is a corpus of approximately 1000 hours of 16kHz read English speech. Our training set was a combination of the train-clean-100, train-clean-360, and train-other-500 subsets. And test-clean is used for evaluation.

For the objective evaluation of our speech codec models, we adopt the methodology proposed by Vocos \cite{siuzdak2023vocos}. We utilize several metrics to assess the performance, including UTMOS \cite{saeki2022utmos} for speech naturalness, PESQ \cite{rix2001perceptual} for perceptual quality, STOI for speech intelligibility, and the F1 score for voiced/unvoiced classification accuracy.
\subsection{Training Setup}
All models are trained on 2 NVIDIA V100 GPUs with a batch size of 10, using 1-second segments randomly cropped from the original utterances. In the first stage, employing the AdamW optimizer with $\beta_1=0.8$ and $\beta_2=0.9$. The learning rate is initialized at 1e-4 and linearly decreases to 1e-5 over 1,000 warmup steps. In the second stage, the batch size is increased to 24, and the learning rate is linearly reduced from 2e-5 to 1e-5. 

And leveraging BigCodec's strong codebook construction, we directly use its official checkpoint for the DS-Codec-VQ's decoder training stage.

\subsection{Comparison Models} \label{baseline}
We evaluate our proposed speech codec by comparing it with several models using their official checkpoints, including Encodec\footnote{https://github.com/facebookresearch/encodec} \cite{defossez2022highfi} (1.5 kbps and 6 kbps), DAC\footnote{https://github.com/descriptinc/descript-audio-codec} \cite{kumar2024high} (4 kbps and 1 kbps), TiCodec\footnote{https://github.com/y-ren16/TiCodec} \cite{ren2024fewer} (with multiple and single codebooks), Wavtokenizer\footnote{https://github.com/jishengpeng/WavTokenizer} \cite{ji2024wavtokenizer}, and BigCodec\footnote{https://github.com/Aria-K-Alethia/BigCodec} \cite{xin2024bigcodec}.
\begin{itemize}
    \item \textbf{Encodec}: A widely recognized speech codec based on Residual Vector Quantization (RVQ) and a mirrored structure. It has been extensively adopted across various domains.
    \item \textbf{DAC}: A state-of-the-art (SOTA) speech codec that integrates RVQ with low-dimensional Vector Quantization (VQ) techniques. With 74M parameters, the model is highly effective.
    \item \textbf{Ticodec}: A neural speech codec designed to separate and independently quantize time-varying and time-invariant information in speech signals.
    \item \textbf{Wavtokenizer}: A discrete acoustic codec model that achieves significant advancements in compression, reconstruction quality, and semantic modeling without employing a mirrored decoder upsampling structure. 
    \item \textbf{Bigcodec}: A low-bitrate neural speech codec operating at 1.04 kbps. The model, scaled to 159M parameters, employs a single codebook with 8,192 codes.
\end{itemize}

Moreover, we also evaluated other speech codecs such as SingleCodec \cite{li2024single}, PQ-VAE \cite{guo2024addressing}, and so on. However, these models have a weak performance on the metrics mentioned in Section \ref{metrics} or only had a small number of test samples. So, we only make a comparison of the above models.

\subsection{Result}

To evaluate the speech reconstruction performance of DS-Codec, we compared it with the models mentioned in Section \ref{baseline}. The objective scores evaluated on the test-clean of LibriSpeech are shown in Table \ref{tab:performance}. 
DS-Codec demonstrates outstanding performance across all objective metrics. Compared to other single-codebook speech codecs, our model exhibits superior capabilities in speech reconstruction tasks, especially under low-bitrate conditions. This performance can be attributed to the effectiveness of the two-stage training framework. In the first stage, the mirror architecture significantly enhances codebook construction and representation learning. In the second stage, the integration of the Transformer Block further improves speech reconstruction quality, allowing the model to focus more effectively on generating high-fidelity speech. 

\begin{table}[ht]
    \caption{The objective metrics scores testing on 2000 utterances selected randomly in LJSpeech.}
    \centering
        \resizebox{\linewidth}{!}{
            \begin{tabular}{cccccc}
            \toprule
            \textbf{Model} & \textbf{UTMOS $\uparrow$} & \textbf{PESQ $\uparrow$} & \textbf{STOI $\uparrow$} & \textbf{F1 Score $\uparrow$} \\
            \midrule
                GT & 4.378 & - & - & - \\
                Wavtokenizer & 3.870 & 1.948 & 0.900 & 0.909 \\
                Bigcodec & 4.385 & 2.822 & 0.951 & 0.943 \\
            \midrule
                DS-Codec-PQ & \textbf{4.428} & \textbf{2.886} & 0.950 & 0.942 \\
                DS-Codec-VQ  & \textbf{4.451} & \textbf{2.962} & \textbf{0.955} & \textbf{0.946} \\
            \bottomrule
        \end{tabular}
        }
\label{tab:out-domain}
\end{table}

To further validate the generalization capability of DS-Codec, we conducted additional evaluations on a subset of the LJSpeech dataset, comprising 2,000 randomly selected utterances. As illustrated in Table \ref{tab:out-domain}, DS-Codec consistently outperforms comparison models across all objective metrics. These results demonstrate the model’s robustness in generalizing to diverse speech data, confirming its effectiveness beyond the training domain.
\subsection{Mirror vs Non-mirror} \label{discussion}
\begin{figure}[t]
  \centering
  
  \includegraphics[width=\linewidth]{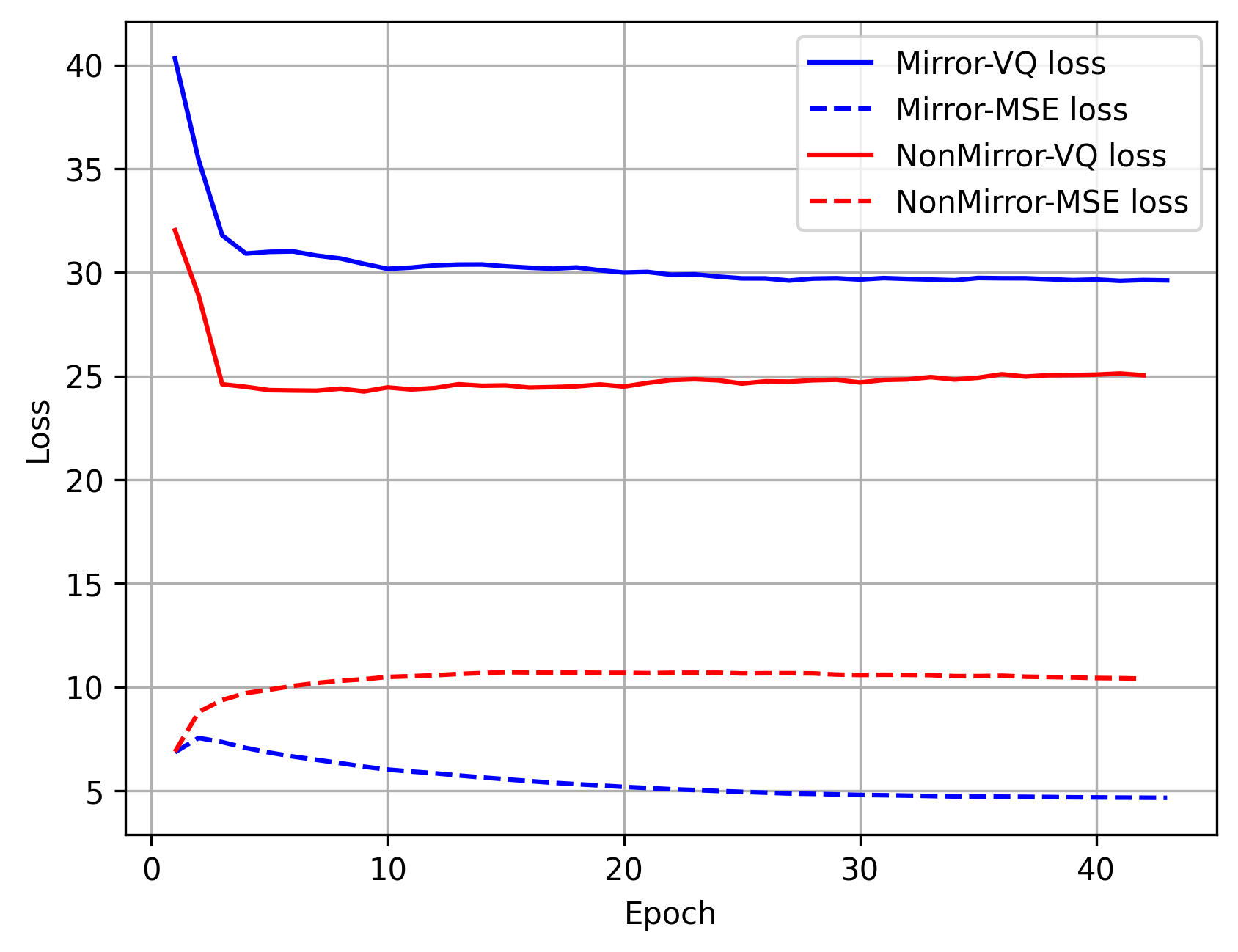}
  \caption{Training Loss Comparison. VQ loss represents the vector quantization loss, while MSE loss is defined as the mean squared error (MSE) loss between the input and output of the Quantization Module. }
  \label{fig:training_loss_comparison}
\end{figure}
We conducted a detailed analysis to evaluate the impact of mirrored and non-mirrored structures on quantization modu during training. Specifically, we monitored the VQ loss and the MSE loss between the quantization module's input and output throughout the training process. The non-mirrored structure was trained using a single-stage joint training approach based on the proposed model architecture, while the mirrored structure was trained following the first stage of our training strategy. As shown in Figure \ref{fig:training_loss_comparison}, the VQ loss reaches convergence after a few training epochs for both architectures. Meanwhile, the MSE loss decreases gradually with increasing training epochs, highlighting the advantage of the mirrored structure. Although the mirrored structure achieves a higher VQ loss compared to the non-mirrored structure, it exhibits a significantly smaller discrepancy between the quantization module's input and output. This demonstrates the mirrored structure's superior codebook construction and representation learning performance.

The primary objective of the Quantization Module is to reduce the discrepancy between the module's input and output. A smaller gap indicates higher fidelity in the reconstructed speech. This concept also can be supported by the results presented in Table \ref{tab:Ablation}. Our proposed Mirror Stage 1 outperforms APCodec+ Stage 1 with a non-mirrored structure across all metrics.
\begin{table}[t]
    \caption{Objective metrics scores of different training strategies based on DS-Codec-PQ testing on the LibriSpeech test set with 2620 utterances. Stage-2-t means without the Transformer Block.}
    \centering
        \resizebox{\linewidth}{!}{
            \begin{tabular}{ccccccc}
            \toprule
            \textbf{Training} & \textbf{Stage}  & \textbf{UTMOS $\uparrow$} & \textbf{PESQ $\uparrow$} & \textbf{STOI $\uparrow$} & \textbf{F1 Score $\uparrow$} \\
            \midrule
            \multirow{2}{*}{APCodec+} 
                    & 1 & 4.113 & 2.632 & 0.931 & 0.939 \\
                    & 2 & 4.186 & 2.754 & 0.937 & 0.940 \\
            \midrule
            \multirow{3}{*}{Proposed}
                & 1 & 4.123 & 2.768 & 0.936 & 0.941 \\
                & 2-t & 4.195 & 2.863 & 0.941 & 0.942 \\
                & 2 & \textbf{4.214} & \textbf{2.882} & \textbf{0.941} & \textbf{0.943} \\
            \bottomrule
        \end{tabular}
        }
\label{tab:Ablation}
\end{table}
\subsection{Ablation Studies}
We conducted ablation experiments to evaluate the effectiveness of the proposed training strategy and compared its performance with the training strategy used in APCodec+ under the same model architecture. In the first stage, APCodec+ employs joint training, which involves a non-mirrored structure where the Encoder, Quantization Module, Transformer Module, and Decoder are trained simultaneously. As shown in Table 3, despite APCodec+ having a stronger decoder for speech reconstruction, its performance in speech reconstruction is inferior to that of the proposed mirror structure, which achieves better results with a weaker decoder. Compared with stage 2-t, the model achieves further improvement after integrating the Transformer Block, underscoring the importance of the decoder in enhancing reconstruction quality. In the second stage of training, the proposed strategy demonstrates significant advantages. It requires fewer training epochs and incurs lower computational costs while achieving substantial improvements in model performance.
This highlights the efficiency and effectiveness of the proposed two-stage training with a mirror-to-non-mirror architecture switching framework.

\section{Conclusion}
This paper introduces DS-Codec, a neural speech codec built on a novel two-stage training strategy. Our model demonstrates superior performance in speech reconstruction by leveraging a two-stage training framework that balances the advantages between mirrored and non-mirrored structures, outperforming previous neural speech codecs. Additionally, we conduct comprehensive experiments to analyze the impact of mirror and non-mirror architectures on model performance. The results show the importance of a robust codebook enabled by the mirrored structure and the critical role of a powerful decoder in enhancing reconstruction quality. In future work, we plan to further optimize DS-Codec and explore its potential to advance research in speech generation tasks.
\section{Acknowledgements}
This work was supported in part by the National Natural Science Foundation of China under Grants 62276220 and 62371407 and the Innovation of Policing Science and Technology, Fujian province (Grant number: 2024Y0068)
\bibliographystyle{IEEEtran}
\bibliography{mybib}

\end{document}